\documentclass[twocolumn,showpacs,preprintnumbers,amsmath,amssymb,APSl,prd,nofootinbib]{revtex4-1}

\usepackage{dcolumn}
\usepackage{bm}
\usepackage{ifpdf}
\usepackage{hyperref}
\usepackage{bm}
\usepackage{xcolor,color,graphicx,graphics}
\usepackage[spanish,english]{babel}
\usepackage[latin1]{inputenc}
\usepackage[OT1]{fontenc}
\usepackage{latexsym,amssymb,amsmath,amsfonts}
\usepackage{makeidx}
\usepackage{epsfig,subfigure}
\usepackage{natbib}
\usepackage{epstopdf}
\usepackage{mathrsfs}
\usepackage{hyperref}
\hypersetup{colorlinks=true, linkcolor=blue, citecolor=green}
\usepackage{enumerate}

\definecolor{red}{rgb}{1,0,0}

\def\+{^\dagger}

\def\<{\leftarrow}
\def\>{\rightarrow}

\def\({\left(}
\def\){\right)}

\newcommand{\bi}{\begin{itemize}} 				\newcommand{\ei}{\end{itemize}}
\newcommand{\benu}{\begin{enumerate}} 		\newcommand{\enu}{\end{enumerate}}
\newcommand{\bd}{\begin{dinglist}{0}}     \newcommand{\ed}{\end{dinglist}}
\newcommand{\bfig}{\begin{figure}[htbp]}  \newcommand{\efig}{\end{figure}}
        			
\newcommand{\bc}{\begin{center}} 				  \newcommand{\ec}{\end{center}}
\newcommand{\be}{\begin{equation}} 				\newcommand{\ee}{\end{equation}}
\newcommand{\bsub}{\begin{subequations}}  \newcommand{\esub}{\end{subequations}}
\newcommand{\ben}{\begin{eqnarray}} 			\newcommand{\een}{\end{eqnarray}}
\newcommand{\ba}[1]{\begin{array}{#1}} 		\newcommand{\ea}{\end{array}}
\newcommand{\bea}{\begin{equation}\begin{array}{rcl}}
\newcommand{\eea}{\end{array}\end{equation}}

\begin{document}

\title{From fundamental physics to tests with compact objects in metric-affine theories of gravity}

\author{Diego Rubiera-Garcia} \email{drubiera@ucm.es}

\vspace{0.8cm}

\affiliation{Departamento de F\'isica Te\'orica and IPARCOS,  Universidad Complutense de Madrid, E-28040 Madrid, Spain}

\begin{abstract}
This work provides a short but comprehensible overview of some relevant aspects of metric-affine theories of gravity in relation to the physics and astrophysics of compact objects. We shall highlight the pertinence of this approach to supersede General Relativity on its strong-field regime, as well as its advantages and some of its difficulties. Moreover, we shall reflect on the present and future opportunities to testing its predictions with  relativistic and non-relativistic stars,  black holes, and other exotic horizonless compact objects.
\end{abstract}




\maketitle
\date{\today}

\section{Introduction}	

\subsection{The strong-field era of gravitational physics}

Einstein's General Theory of Relativity (GR) is alive and well. Already one hundred years after the expeditions to Principe Island  (Eddington and Cottingham) and Sobral (Crommelin and Davidson) \cite{Crispino:2019yew} to test Sun's light deflection, we have accumulated plenty of evidence from solar system experiments, post-Newtonian tests, gravitational lensing, test on the equivalence principle, frame-dragging effects, etc, on the reliability  of this theory to describe gravitational phenomena \cite{Will}. Moreover, the theory has been built in the cosmological concordance $\Lambda$CDM model, which has successfully met all observations at small and large scales \cite{Bull:2015stt}. In addition, we have witnessed the beginning of a new era on astrophysics of compact objects following the discovery of gravitational waves, consistently interpreted as the coalescence of two compact objects: black hole - black hole \cite{Abbott:2016blz}, and black hole - neutron star \cite{TheLIGOScientific:2017qsa}. Tests on the Kerr black hole hypothesis itself have been performed according to the radiation emitted from the accretion disks surrounding them \cite{Bambi:2011mj}, as well as by the measurement of the shadow of the supermassive central object of the M87 galaxy \cite{Akiyama:2019cqa}. In all these observations GR has fully met if not surpassed our expectations. This has been possible thanks to more than half a century of powerful theoretical developments, by the huge improvement in the capabilities of numerical relativity, and by the support received from large international collaborations. This has triggered the beginning of a new era where the possibility of testing the strong-field regime of the gravitational interaction is at hand. But will new Physics be found?. What could we expect?.

\subsection{The need to go beyond GR}

If GR is a successful theory, why the need to going beyond it?. First of all, the $\Lambda$CDM model requires the introduction of extra matter fields (inflation, dark matter and dark energy) with unusual properties and which, despite intense and varied observational searches \cite{Bertone:2004pz}, has not been directly detected at any terrestrial experiment yet. Moreover, the tension in the value of the Hubble constant as given by the discrepancies from the direct local measurements and the model-dependence inference from CMB data continues to puzzle cosmologists \cite{Bernal:2016gxb}.  On a more fundamental level, the well known incompatibility between GR and quantum mechanics has been a powerful drive for decades to search for an hypothetical quantum theory of gravity superseding GR \cite{Oriti}, though with relatively little success. Moreover, GR is prone to the existence of space-time singularities, which unavoidably arise in the innermost region of black holes and in the early Universe \cite{Senovilla:2014gza}. From the astrophysics of compact objects, we have the challenge of generating neutron stars above two solar masses with realistic equations of state to meet observations \cite{Antoniadis:2013pzd}, a problem that could worsen with time as even heavier neutron stars are detected. Other issues of interest are the recent suggestion about the existence of ``Super-Chandrasekhar" white dwarfs with masses in the range $2-2.8M_{\odot}$ \cite{Howell:2006vn}, which would defy the standard picture of stellar evolution, and the potential existence of new (exotic) compact objects with different properties than the Kerr solution, whose existence could be revealed thanks to gravitational waves \cite{Mark:2017dnq}.

Within this context, ``modified gravity" becomes a buzzword for many proposals to extending GR following different prescriptions. The corresponding literature is exceedingly large \cite{DeFelice:2010aj,CLreview,Clifton:2011jh,Nojiri:2017ncd,Heisenberg:2018vsk}, and a bunch of predictions have been developed within  astrophysical and cosmological scenarios. Nowadays, many such proposals (mostly those motivated by cosmological considerations) are heavily constrained by the  observation $c_{GW}=c$ (up to a $\sim 10^{-15}$ precision) by the LIGO/VIRGO Collaboration, as discussed in Ref.\cite{EzZuma}.  In most such extensions of GR, the metric tensor is regarded as the only player in town (metric approach), while the connection is \emph{violently imposed} \cite{Zanelli:2005sa} to be given by the Christoffel symbols of the metric (that is, the Levi-Civita connection).  This ad hoc constraint on the nature of the connection has been inherited from traditional/educational reasons on the way GR is usually seen and thought, and has been consequently propagated through most of the literature in the field. However, there are alternatives to this paradigm.

\subsection{The role of the affine connection}

It is indeed known by mathematicians since a very long time ago that, in general, any  affine connection $\Gamma \equiv \Gamma_{\mu\nu}^{\lambda}$ can be decomposed into its curvature (associated to a Levi-Civita connection), torsion (associated to its antisymmetric part) and non-metricity (associated to the failure of the connection to yield $\nabla^{\Gamma}_{\alpha} g^{\beta \gamma}=0$) pieces \cite{Eisenhart,Schouten,Hehl:1994ue}. We are used to think on GR as the (Riemannian) theory of gravity where torsion and non-metricity are set to zero, while we build the lowest-order action on scalar objects made up of the curvature piece (that is, the Einstein-Hilbert action). However, in what has been recently popularized as \emph{the geometrical trinity of gravity} \cite{BeltranJimenez:2019tjy}, we know now that there exists three equivalent (modulo some technicalities regarding boundary terms) formulations of GR. The first alternative formulation to the standard curvature-based one is the teleparallel equivalent of GR \cite{Maluf:2013gaa}, in which we switch off curvature and non-metricity, but keep torsion. On the other hand, in the symmetric (or coincident) teleparallel GR \cite{Nester:1998mp} we switch off curvature and torsion, but keep non-metricity. The corresponding theories succeed in yielding the same observational predictions as those of (curvature-based) GR when the lowest-order scalar objects are considered in the action. This result puts forward the richness encoded in the affine connection, which therefore could play a more fundamental role in the gravitational dynamics that thought in the past. Moreover, this observation offers a new landscape of possibilities for extending GR depending on the assumptions made upon these three pieces of the affine connection. In this work we shall focus on a particular formulation of modified gravity which, for the sake of this paper, shall be dubbed as metric-affine theories of gravity, and discuss the open opportunities to test the predictions of these theories within the astrophysics of compact objects.

\section{Metric-affine formulation of theories of gravity}

By metric-affine gravity we mean those theories where metric and connection are regarded as independent degrees of freedom \cite{Olmo:2011uz}. Current research has identified a promising family of such theories to be theoretically and observationally viable, dubbed as  Ricci-based gravities (RBG), and given by the action
\begin{equation} \label{eq:SG}
S=\frac{1}{2\kappa^2} \int d^4x \sqrt{-g} \mathcal{L}_G[g_{\mu\nu},R_{\mu\nu}(\Gamma)]+S_{m}[g_{\mu\nu},\psi_m] \ ,
\end{equation}
where $\kappa^2$ is Newton's constant in suitable units, and $g$ is the determinant of the space-time metric $g_{\mu\nu}$. In order for this action to be a scalar, the dependence on the geometrical objects in the gravitational Lagrangian $\mathcal{L}_G$ must appear in terms of powers of traces of the object ${M^\mu}_{\nu}\equiv g^{\mu\alpha}R_{\alpha\nu}$, where $R_{\mu\nu}(\Gamma)$ is the symmetric part of the Ricci tensor (which is a priori independent of any metric). In this version of metric-affine gravities, torsion can be safely set to zero since in the matter sector $\mathcal{S}_m$ we are assuming that the matter fields $\psi_m$ couple to the metric, but not to the connection \cite{Afonso:2017bxr} (which would be relevant, for instance, should one consider fermions in the matter sector), in order to comply with the equivalence principle. The requirement of a symmetric Ricci tensor is justified on the grounds that the consideration of an antisymmetric piece would make the theory bump into troubles with ghost-like instabilities \cite{BeltranJimenez:2019acz}. This family of actions is wide enough to cover many interesting cases in the literature such as GR itself, $f(R)$ theories, quadratic gravity, Born-Infeld inspired theories of gravity, and so on. It is worth pointing out that, should the action be formulated \emph{\'a la metric}, that is, by imposing the metric-connection compatibility condition, $\nabla_{\mu}^{\Gamma} (\sqrt{-g} g^{\alpha \beta})=0$, then one would immediately get lost into troubles with higher-order field equations, ghost-like instabilities,  incompatibility with solar system experiments, etc, though some restrictions on $\mathcal{L}_G$ may alleviate some of these problems.

Let us consider the metric-affine formulation of this family of theories. When $g_{\mu\nu}$ and $\Gamma^\alpha_{\beta\gamma}$ are independent then the equations of motion obtained from the variation of the action (\ref{eq:SG}) with respect to both of them can be cast under the Einstein-like form \cite{Belt,Delhom:2019zrb}
\begin{equation}\label{eq:GmnGeneral}
{G^\mu}_\nu(q)=\frac{\kappa^2}{|\hat{\Omega}|^{1/2}}\left[{T^\mu}_\nu-{\delta^\mu}_\nu\left(L_G+\tfrac{T}{2}\right)\right] \ ,
\end{equation}
where $T$ is the trace of the stress-energy tensor ${T^\mu}_\nu=\frac{-2}{\sqrt{-g}}\frac{\delta \mathcal{S}_m}{\delta g^{\mu\nu}}$, while ${G^\mu}_\nu(q)$ is the Einstein tensor of a new metric $q_{\mu\nu}$ satisfying $\nabla_{\mu}^{\Gamma} (\sqrt{-q} q^{\alpha \beta})=0$ (so that the independent connection can be obtained as the Christoffel symbols of $q_{\mu\nu}$), which is related to the space-time metric as
\begin{equation} \label{eq:qggen}
q_{\mu\nu}=g_{\mu\alpha}{\Omega^\alpha}_{\nu} \ .
\end{equation}
The \emph{deformation matrix} ${\Omega^\alpha}_{\nu}$ depends on the particular $\mathcal{L}_G$ chosen, but \textit{it can always be written on-shell} as a function of the stress-energy tensor, ${T^\mu}_\nu$. For instance, in the $f(R)$ case the relation above becomes conformal: $q_{\mu\nu}=f_R(T) g_{\mu\nu}$, while for other RBGs this relation will be a full algebraic transformation involving the mixture of all components of the deformation matrix with the space-time metric. Note that in GR, $\mathcal{L}_G=R$, one has $G_{\mu\nu}(q)=\kappa^2 T_{\mu\nu}$ (perhaps supplemented with a $\Lambda$ term) and $q_{\mu\nu}=g_{\mu\nu}$ (modulo a trivial re-scaling) and thus the metric-affine formulation of GR yields exactly the same dynamics and predictions as the standard one\footnote{For a recent discussion on the interpretation of certain geometrical properties of specific solutions in both formalisms see Ref.\cite{Bejarano:2019zco}.}.

The RBG family  (or, at least, most of its members, such as those modifying GR only in the ultraviolet limit) enjoys a number of distinctive and physically appealing features:

\begin{itemize}

\item Second-order field equations.

\item Vacuum solutions are those of GR.

\item No ghost-like instabilities.

\item $c_{GW}=c$ and two tensorial polarizations.

\end{itemize}

The above properties ensure the consistence of (most) RBGs with solar system experiments and with gravitational wave observations so far. The Einstein-field representation (\ref{eq:GmnGeneral}) clearly shows that these theories in their dynamics for $q_{\mu\nu}$ can be interpreted as GR with new matter couplings engendered via both the deformation matrix ${\Omega^\alpha}_{\beta}({T^\mu}_\nu)$ and the gravitational Lagrangian $\mathcal{L}_G({T^\mu}_\nu)$. This observation shall be of great relevance later. Moreover, due to this, the trademark of RBGs is that the new physical effects will be fed by the \textit{energy density} of the matter fields, and not just by the integration over sources. This has important consequences both for fundamental physics and for the expectations regarding the properties of compact objects, and provides a fertile playground to explore new physics beyond GR.

\section{Theoretical physics of black holes}

The Einstein-like representation of the field equations (\ref{eq:GmnGeneral}) has allowed to introduce suitable methods for the sake of finding explicit solutions in different contexts, which has triggered a quick progress in the understanding of black holes and other compact objects within these theories. Extensions of these theories and methods for other cases, such as the role of the Riemann tensor or the presence of torsion, has also begun to be unravelled. In this section, we shall highlight some theoretical findings regarding black hole physics within metric-affine/RBG theories, which shall pave the path to make contact with the astrophysics of compact objects.

\subsection{Spherically symmetric black holes}

Spherically symmetric black holes have been the most frequent playground to test the predictions of these theories for compact objects, thanks to the possibility of finding solutions under analytic form corresponding to different matter sources, which simplifies their analysis. The simplest models at this regard are $f(R)$ theories, since the relation between the two metrics becomes conformal. However, in this case, the trace of the corresponding field equations yields $Rf_R-2f=\kappa^2 T$, telling us that $R=R(T)$. This implies that the dynamics encoded in the new  contributions to the field equations (\ref{eq:GmnGeneral}) can only be excited in presence of matter-energy sources with a non-vanishing trace. This result prevents using Maxwell electrodynamics to find the counterpart of the Reissner-Nordstr\"om solution of GR, and forces one to use non-linear electrodynamics instead \cite{Olmo:2011ja}.

For more general RBGs, however, the full stress-energy tensor will appear in the dynamics of the theory. In such cases, the main difficulty to be sorted out is to resolve Eq.(\ref{eq:qggen}) to find the relation between curvature and stress-energy tensor, which requires an analysis case-by-case. For instance, in the case of quadratic gravity, $\mathcal{L}_G=R+aR^2 + bR_{\mu\nu}R^{\mu\nu}$, with $a$ and $b$ some parameters, this relation can be explicitly found for spherically symmetric solutions after some algebra \cite{Olmo:2012nx}. A particularly interesting theory at this regard is the so-called Eddington-inspired Born-Infeld gravity (EiBI) \cite{BeltranJimenez:2017doy}, which is given by the action
\begin{equation} \label{eq:EiBI}
\mathcal{S}_{EiBI}=\frac{1}{2\kappa^2} \int d^4x \left(\sqrt{-\vert g_{\mu\nu}+\epsilon R_{\mu\nu} \vert} - \lambda \sqrt{-g} \right) \ ,
\end{equation}
where $\epsilon$ is a parameter with dimensions of length squared. Moreover, the theory features an effective cosmological constant given by $\Lambda=\frac{\lambda-1}{\kappa^2 \epsilon}$. In this case,
the expression for the deformation matrix is remarkably simple,
\begin{equation}
\vert \Omega \vert^{1/2} [{\Omega^{-1}]^\mu}_{\nu}=\lambda \delta^{\mu}_{\nu} - \epsilon \kappa^2 {T^\mu}_{\nu} \ ,
\end{equation}
which can be explicitly solved for a given ${T^\mu}_{\nu}$ via an ansatz for the deformation matrix mimicking its algebraic structure (plus a diagonal term if not present). For some matter sources such as electromagnetic fields \cite{Olmo:2013gqa} and, more generally, some types of anisotropic fluids \cite{Menchon:2017qed}, this strategy allows to find exact black hole solutions out of the RBG field equations (\ref{eq:GmnGeneral}) following pretty much the same procedure for their resolution as in the GR case. A key aspect in this analysis is to set two line elements for the $g_{\mu\nu}$ and $q_{\mu\nu}$ geometries respecting the symmetries of the problem (spherical symmetry in this case), while working out explicitly the relation (\ref{eq:qggen}) between the metric functions in both frames. This procedure is quite efficient, and allows to circumvent any need to solve the (highly complicated) structure of the RBG field equations should they be written directly in terms of the $g_{\mu\nu}$ geometry.

The final conclusion of this analysis is that, for all RBGs studied so far with the matter fields above, the line element of any static, spherically symmetric solution can be conveniently cast under the form
\begin{equation} \label{eq:linesss}
ds^2=-\frac{A(x)}{\Omega_1(x)}dt^2+ \frac{dx^2}{A(x)\Omega_2(x)}+r^2(x) d\Omega^2 \ ,
\end{equation}
where $d\Omega^2=d\theta^2 + \sin^2(\theta)d\varphi^2$ is the volume element in the unit two-sphere, while the functions $\Omega_1(x),\Omega_2(x)$ characterize the particular combination of RBG + matter field description, and typically contain the mass and charge of the solution as well as additional parameters coming from the RBG Lagrangian density. Moreover, if the RBG theory modifies GR in the strong-field regime (like in the case of EiBI gravity (\ref{eq:EiBI})), then the corresponding solutions will boil down to the Reissner-Nordstr\"om one of GR in their weak-field limit.

 The function $A(x)$ in (\ref{eq:linesss}) encodes the modified description of horizons. Due to the fact that high-enough local energy densities are typically attained only in the innermost region of black holes, the effects of RBGs manifest also there, while presumably leaving only very tiny imprints on the region outside the event horizon\footnote{The case with scalar fields is an exception to this general rule \cite{Afonso:2019fzv}, which shall be discussed in Sec.\ref{sec:scalar}.}. Despite this, the general structure of horizons may undergo large modifications, finding, in addition to the standard two, single (degenerate) or none horizons of the Reissner-Nordst\"om solution of GR, configurations with a single (non-degenerate) horizon (thus bearing a closer resemblance to the Schwarzschild solution instead), or solutions where the metric functions are finite at the center. This structure of horizons mimics the one found in certain models of non-linear electrodynamics in the context of GR \cite{Fernando:2003tz}.

As for the radial function $r^2(x)$, for matter-energy sources whose stress-energy tensor can be expressed as
\begin{equation} \label{eq:Tmunuflu}
{T^\mu}_{\nu}=\text{diag}(-\rho,-\rho,K(\rho),K(\rho)) \ ,
\end{equation}
it is given by \cite{Nascimento:2019qor}
\begin{equation}
r^2(x)=\frac{x^2}{\Omega_2(x)} \ .
\end{equation}
It is worth stressing that this functions does not need to be monotonic. When this happens, this radial function is capable to yield a bounce at some $x=x_c$ ($r=r_c$), which can be interpreted in some cases as a signal of a wormhole structure, which typically allows for the extensibility of geodesics beyond $r=r_c$ \cite{Visser}, as discussed in next section. Though wormholes unavoidably violate standard energy conditions within the context of GR, this is not necessarily so within RBGs thanks to the extra gravitational corrections, that can be understood as yielding an effective stress-energy tensor. Let us also note that is possible to introduce new coordinates to rewrite the line element (\ref{eq:linesss}) in the more canonical form $ds^2=-\tilde{A}(y)dt^2+\tilde{A}^{-1}(y)dy^2 +r^2(y)d\Omega^2$ (so therefore the contributions of $\Omega_1,\Omega_2$ would be hidden within $\tilde{A}$ and the radial coordinate $y$) though this change usually spoils any explicit simple representation of the radial function $r^2(y)$.

\subsection{Regular black holes}

The theorems on singularities developed by Penrose and Hawking, among others \cite{Theo1,Theo2,Theo3,Theo4}, tell us that GR is prone to the existence of incomplete causal geodesic curves in the manifold. As null geodesic curves are associated to the paths of light rays and time-like geodesics to the free-falling of physical observers, the existence of any such curve would imply the breakdown of the predictability of GR. This unavoidably happens, for instance, deep inside black holes and in the Big Bang singularity. Therefore, \textit{geodesic completeness} nicely captures the intuitive idea that in a physically reasonable space-times observers or information should not suddenly cease to exist or to emerge from nowhere \cite{Curiel}, and has become the main criterion in the literature to classify regular/singular space-times.

The gravitational community has engaged for decades in the search for black hole solutions overcoming such  singularities, yielding a fruitful field of research dubbed as \emph{regular black holes}. To build such solutions one has to remove any of the hypothesis of the singularity theorems, which in one of their canonical formulations read \cite{Senovilla:2014gza,Senovilla:2018aav}

\begin{itemize}

\item A future trapped surface is developed.

\item Fulfilment of the null congruence condition (equivalent to the fulfilment of the null energy condition via Einstein equations).

\item Global hyperbolicity.

\end{itemize}

These three conditions guarantee the existence of a focusing point preventing the continuation of the wordline of every observer. Unsurprisingly, the literature on this field has truly blossomed \cite{Ansoldi:2008jw}, with quite a fair collection of such regular black hole solutions removing any of these hypothesis. In this quest, most attempts have focused on finding black holes whose curvature scalars are everywhere regular rooted on the fact that\footnote{Curvature scalars are also typically easier to characterize;  in particular they can be easily programmed via dedicated tools in Mathematica, in order to determine the matter sources able to yield finite such scalars.}, though the singularity theorems speak nothing on the behaviour of curvature scalars, almost every geodesically incomplete solution ever found in GR has also divergent curvature scalars, with a few exceptions \cite{SKMHHBook}.

\begin{figure*}[t!]
\centering
\includegraphics[width=6.2cm,height=5cm]{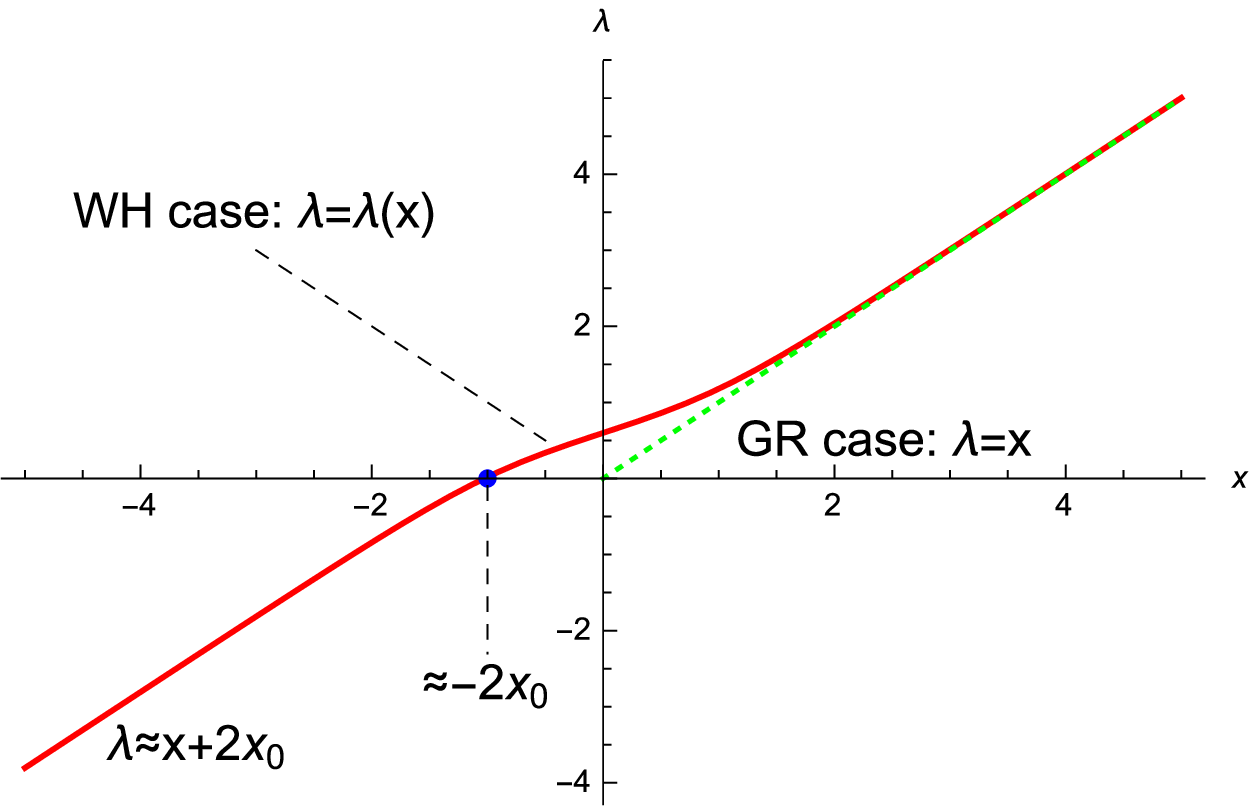}
\includegraphics[width=6.2cm,height=5cm]{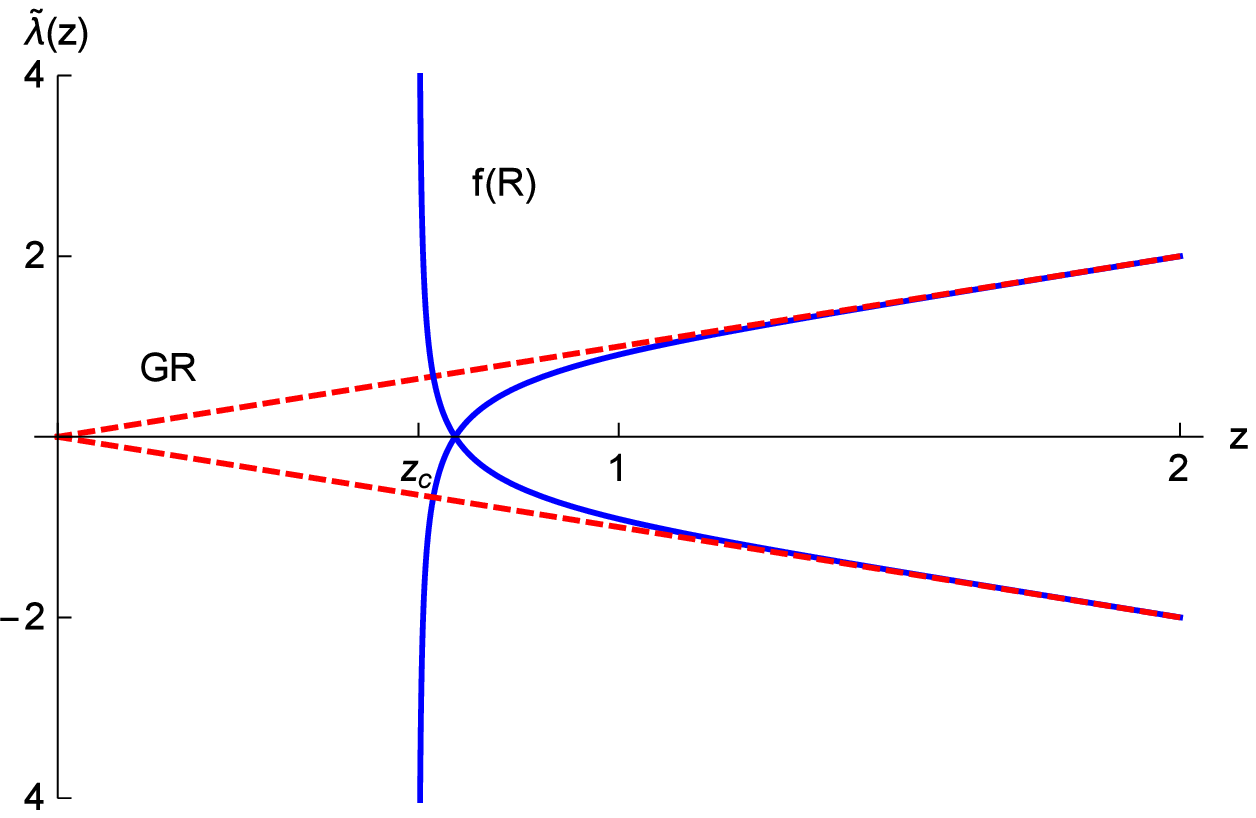}
\caption{Two different mechanisms for the extensibility of the affine parameter at $x=r_c$ (or $z \equiv r/r_c=1$ in dimensionless variables) for null radial geodesics. Left figure: via a bounce in the radial function (using quadratic gravity coupled to Maxwell electrodynamics), extracted from Ref.\cite{Olmo:2015bya}. Right figure: the central region lies on the future (or past) boundary of the manifold, requiring an infinite affine time to reach it (using $f(R)=R+\alpha R^2$ coupled to Born-Infeld electrodynamics), extracted from Ref.\cite{Bambi:2015zch}. The straight line(s) $\lambda=\pm x$ in both cases corresponds to the incomplete null radial geodesics of the Reissner-Nordstr\"om solution of GR.}
\label{fig.1}
\end{figure*}

Metric-affine gravities have their share of regular black holes \cite{Olmo:2016tra}. As with many other regular black holes in modified theories of gravity, the fact that the field equations are different from the Einstein equations allows for the second of the conditions above to relate in a different way the focusing of geodesics and the fulfilment of the energy conditions. In other words, there is the possibility for some effective stress-energy tensor sourcing the new set of generalized Einstein equations to violate the focusing condition, but such that the physical stress-energy tensor (the one derived from the matter action (\ref{eq:SG})) satisfies the energy conditions. Regardless of these considerations, for a stress-energy tensor of the form (\ref{eq:Tmunuflu}) one can write the geodesic equation  for the line element (\ref{eq:linesss}) as
\begin{equation} \label{eq:geo}
\frac{1}{\Omega_1^2}\left(\frac{dx}{d\lambda}\right)^2=E^2-\frac{A(x)}{\Omega_1}\left(\frac{L^2}{r^2(x)}- k \right) \ ,
\end{equation}
where $\lambda$ is the affine parameter (the proper time for a time-like observer), $k=0,-1$ for null and time-like geodesics, and $E,L$ are the energy and angular momentum per unit mass, respectively. From a conceptual point of view one can envisage two basic mechanisms for Eq.(\ref{eq:geo}) to yield complete geodesics: i) either some \emph{bounce} arises in $r(x)$ near that region where the point-like singularity should be, $x=x_c$, allowing geodesics to defocus and continue their path to another region of space-time, or ii) the central region is displaced in such a way that every (null and time-like) geodesic takes an infinite time to reach to it (see Refs.\cite{Carballo-Rubio:2019nel,Carballo-Rubio:2019fnb} for an extended discussion on these two mechanisms and their interpretation and consequences).

 In Fig.\ref{fig.1} we depict explicit implementations of both mechanisms within RBGs coupled to electromagnetic fields, where null radial geodesics (which are incomplete in the Reissner-Nordstr\"om geometry of GR) turn out to be complete. Similarly, one can verify that in both cases every other null and time-like geodesic sees an effective potential such that in case i) those geodesics able to overcome the potential barrier can also cross the bouncing region $x=x_c$ ($r=r_c$) and expand to another region of space-time, while in case ii) they require an infinite amount of energy to overcome it and get to $x \rightarrow - \infty$. This guarantees the null and time-like geodesic completeness in these two classes of space-times.

In the latter case one does not need to worry about pathologies in the behaviour of curvature scalars there, since no geodesic will be able to interact with such regions, not even observers with arbitrary (but bounded) acceleration \cite{Olmo:2017fbc}. In the former case one would be tempted to associate this defocusing phenomenon to the curvature of space-time being regularized somehow.  This is not necessarily true in metric-affine gravities. Indeed, there is by now evidence of a breakdown in the correlation between (in)completeness of geodesics and the existence of divergences in (some) curvature scalars. Table I of Ref.\cite{Bejarano:2017fgz} (which corresponds to quadratic $f(R)$ gravity coupled to an anisotropic fluid satisfying standard energy conditions) clearly shows this: neither curvature divergences prevent the extensibility of geodesics, nor the existence of an incomplete geodesic is triggered by infinite curvature. At most, the only correlation found there is that the presence of infinite energy densities do imply incompleteness of geodesics. This last statement could be related to the specific role played by the local energy densities to trigger the new dynamics encoded in RBGs.

What is then the impact of divergent curvature?. One might indeed worry that, no matter what the behaviour of geodesics might be, a physical (extended) observer passing through a divergent-curvature region surely would undergone any utterly disruptive process. From the criteria widely employed in the literature  \cite{CK85} this will be so when the divergence is strong enough so as for the volume element of an extended observer to shrink to zero, as it happens, for instance, in the standard Schwarzschild solution of GR. This issue is still a matter of controversy, with some cases running away from simple interpretations \cite{Ori:2000fi}. To overcome such difficulties, another idea recently brought forward in the literature to look for possible pathologies is to study the interactions which bind together any extended body. For the latter to hold such interactions must be of course strictly causal. This way, in geodesically complete space-times with divergent curvature scalars one would need to study the propagation of light rays from one part or the body to another to determine whether it would be unavoidably destroyed or not, finding that this is not necessarily the case, and that observers could actually survive the trip across such regions \cite{Olmo:2016fuc}.

\subsection{Dynamical scenarios}

The replacement of point-like singularities of (spherically symmetric) black holes in terms of extended structures allowing for the bounce of geodesics raises new conceptual problems. Indeed, where the GR manifold is single-connected, in these new geometries we have structures with non-trivial topologies, so we have to face the always problematic issue of topology change \cite{Horowitz:1990qb}. One might think that these are highly idealized scenarios, where the condition of staticity allows one to play tricks to obtain a desired results. It is therefore of relevance to study whether such geometries generated dynamically, that is, via gravitational collapse \cite{JoshiBook}, can lead to the desired result of the generations of finite-size structures in the central region of black holes. This has been investigated using simplified dynamical scenarios where either vacuum or a pre-existing black hole is sourced by a flux of particles carrying mass and charge with large enough intensity (Vaidya-type solutions), finding that this is indeed the case \cite{Lobo:2014zla}. Such fluxes open up an evolving finite-size structure which relaxes into the static throat once the flux is over, and allows for the completeness of geodesics.

One can look for further insights on this phenomenon by borrowing an analogy with well studied laboratory systems. Indeed, in the solid state physics field of crystalline structures (which have a regular pattern arranging its microscopic constituents) the existence of different types of defects are very well known and which, rather than inducing any pathology are essential in the generation of macroscopic (collective) properties of the material, such as viscosity, plasticity, etc. \cite{KleinertBook}. What is perhaps less known is that such materials admit (actually \emph{require}) a geometry of metric-affine type for their proper description in the continuum (macroscopic) regime, with deep implications for the interpretation of space-time singularities and their relation with specific geometries \cite{Lobo:2014nwa}.

In this section we have illustrated with the case of spherically symmetric black holes how metric-affine theories of gravity offer an interesting playground to test modifications of GR on its strong-field regime. Next we shall study some phenomenological aspects of interest attached to the astrophysics of compact objects within these theories. For a broad overview of the state-of-the-art of this field regarding compact stars see the recent review \cite{Olmo:2019flu}.

\section{Tests with compact objects}

\subsection{Relativistic stars}

Thanks to the quick technological progress achieved in the last few decades, the field of compact stars has seen a great leap in our understanding of the span covered by neutron stars masses and radii \cite{Ozel:2016oaf}. The main conclusion is that neutron stars' radii typically lie between $\sim 10-14$km, and that they can be as massive as $2M_{\odot}$ \cite{Antoniadis:2013pzd} and possibly even more. These improvements in our capabilities to measure the properties of these objects has sparked a renewed interest in testing the predictions of both GR and modified gravity against the phenomenology of neutron stars. A fundamental and long-standing difficulty to tackle this challenge is the fact that at the densities reached at the neutron star's center (up to $5-10$ times the nuclear saturation density), the equation of state (EoS) relating energy density and pressure, $P =P(\rho)$, is unknown. Therefore, obtaining any such EoS requires extrapolations from nuclear physics knowledge using QCD, effective models, etc. This information is needed in order to feed the Tolman-Oppenheimer-Volkoff (TOV) equations \cite{TOV1,TOV2}, describing spherically symmetric stars in hydrostatic equilibrium. As hundreds of EoS exist in the market following different prescriptions (see e.g. Ref.\cite{YagNun} for details), such predictions are highly degenerate.  Introducing metric-affine gravities into the game further complicates things, since every such theory typically carries an additional parameter, and it becomes a hard challenge to extract clear and clean observational discriminators against GR predictions.

For spherically symmetric stars the main relevant outcome of numerical simulations aimed to resolve the TOV equations based on a given EoS, once an RBG theory is selected, is the mass-radius relation since it can be directly confronted with observations by tracking enough neutron stars \cite{vitor,delsate,harko,qauli,pannia}. In particular, the compatibility of the maximum allowed mass with the $2M_{\odot}$ threshold becomes a direct test on the viability of any such scenario (combined with a given EoS). For rotating (slowly, rapidly and differentially) stars, the moment of inertia is another sensible quantity which can be measured, though research in this context is quite scarce \cite{panidel}, as opposed to the metric formalism. In the latter case, the deviations triggered by modified gravity are larger in the moment of inertia than in the mass-radius relations, thus suggesting a better opportunity using this feature to testing the predictions of metric-affine theories as well.

Though the TOV equations for many metric-affine gravities are known,  specific predictions require a case-by-case analysis, which is met with some technical difficulties for specific models. In particular, special care is required to handle the matching to an external (vacuum) solution, since discontinuities in density profiles may be ill-defined in metric-affine gravity. The so-called \emph{surface singularities} \cite{pani} arise precisely from an attempt to employ matched GR solutions on the metric-affine side, which induces the emergence of local divergences in curvature scalars when certain polytropic EoS of physical interest are employed. Therefore, simple GR models seem not to have counterpart on the metric-affine side (the same happens in metric $f(R)$ gravity \cite{Senovilla:2013vra}) requiring both a upgrade of the junction conditions at the stellar surface \cite{OlRuPaper} and a consideration of dynamical aspects (atmospheres, thermodynamics, radiation fluxes, etc) to correctly model stellar surfaces in metric-affine gravity.

\subsection{Non-relativistic stars}

For non-relativistic stars, $P \ll \rho$, the TOV equations can be reduced to their Newtonian counterparts. The relevance of this non-relativistic limit is that white, brown, and red dwarfs can be well modelled in this regime by  polytropic EoS, namely:
\begin{equation} \label{eq:polyeos}
P=K\rho^{ \frac{n+1}{n}} \ ,
\end{equation}
where $K$ is the polytropic constant and $n$ the polytropic index. The corresponding modified (Poisson) equation will typically arise as a number of terms correcting the Lane-Emden equation of GR \cite{GlenBook}
\begin{equation}
\frac{1}{\xi^2} \frac{d}{d\xi} \left( \xi^2 \frac{d\theta}{d\xi}\right)+\theta^n=0 \ ,
\end{equation}
where $\xi$ and $\theta$ are the radial coordinate and the density in suitable rescaled coordinates, respectively. In both GR and on its modified metric-affine version, the zeros of the function $\theta(\xi)$ allow to find the star's masses and radii. The effect of the new RBG corrections is to yield a strengthening/weakening of the gravitational interaction inside astrophysical bodies, with a large impact on many of the properties of such stars. Indeed, since non-relativistic stars are more weakly dependent on unknown non-gravitational elements than their relativistic counterparts, they offer a cleaner scenario in putting to experimental test specific  predictions of RBGs. Let us illustrate this with some examples.

Brown dwarfs encompass a large family of objects with different chemical properties and evolutions \cite{Burrows}, spanning the range of masses between Jupiterian planets (low-mass brown dwarfs) to substellar objects lying at the bottom of the main sequence  (high-mass brown dwarfs). It is precisely at these two limits where ideal scenarios for testing the predictions of metric-affine gravities are found. For high-mass brown dwarfs, which can be modelled with $n=3/2$ in Eq.(\ref{eq:polyeos}) \cite{KippBook},  GR yields an analytic estimate of $M_{MMSM}\sim 0.09 M_{\odot}$  for the minimum mass required for a star to burn sufficiently stable hydrogen to compensate photospheric losses\footnote{When other elements are included in this description, such as thermodynamics, modelling of atmospheres, luminosities, heat transfer, etc, the numerical simulations decrease this bound by a $\sim 10 \%$.}. The same computations can be done within the context of RBG, for instance, in quadratic (Starobinsky) $f(R)$ gravity $f(R)=R+ \beta R^2$. Indeed, an explicit formula for the $M_{MMSM}$ can be obtained in this case depending on $\alpha \equiv \kappa^2 c^2 \beta \rho_c$ \cite{Olmo:2019qsj}, where we see again the dependence of the new dynamics of metric-affine gravities on the local energy densities, this time via the star's central density $\rho_c$. The power of this scenario is clearly seen from the fact that the branch $\alpha>0$ leads to an strengthening of the gravitational interaction allowing for larger minimum masses and, indeed, for $\alpha \gtrsim 0.010$ this $M_{MMSM}$ limit becomes comparable to $M=(0.0930\pm0.0008) M_\odot$, corresponding to the M-dwarf star G1 866C \cite{Segransan}, which is the lowest main-sequence star mass ever observed. Therefore, significantly higher values than this one would presumably be in conflict with observations. On the branch $\alpha <0$ this effect is reversed and compatibility with current observations is guaranteed.

Regarding low-mass brown dwarfs, there is another feature suitable for observational purposes: the minimum mass required for deuterium-burning, which marks the minimum mass limit of a brown dwarf (described  by $n=1$ in Eq.(\ref{eq:polyeos})). In GR this value is given by $M_{MMDB}\sim 0.011-0.016M_{\odot}$ \cite{Spiegel} but, as it happens with the minimum hydrogen burning mass limit, it significantly depends on the assumptions upon the internal composition or the metallicity. This way, one can track the predictions of RBGs for this limit too, since it can be confronted with observations. For instance, as shown in Ref.\cite{Rosyadi:2019hdb} in the case of EiBI gravity, the combination of both limiting masses $M_{MMSM}$ and $M_{MMDB}$ via statistical analysis allows to constraint EiBI parameter as $-1.59 \times 10^2 \leq \epsilon \leq 1.16 \times 10^2$m$^5$kg$^{-1}$s$^{-2}$. Finally, the radius plateu (the constancy of the star's radius with the mass) of low-mass brown dwarfs could be also another test for the predictions of these theories \cite{Sakstein:2015aac}.

White dwarfs, arising from fuel-exhausted main-sequence stars where the gravitational collapse is halted by electron's degeneracy pressure, also offer suitable scenarios to test the predictions of RBGs. For instance, tests on the Chandrasekhar's $1.44M_{\odot}$ limit can be performed \cite{Reijonen:2009hi}, while the suggestion in the literature on the existence of super-Chandrasekhar stars, with masses up to $2.8M_{\odot}$ \cite{Howell:2006vn} has not been addressed within RBGs yet.

\subsection{Rotating black holes and the mapping method}

Real black holes in the Universe do rotate. While a tiny amount of charge is expected to be retained by such objects, it can be completely disregarded in astrophysically sensible scenarios, reducing the Kerr-Newman family to the simpler Kerr one. That the Kerr solution of GR can be reliably used to describe black holes has been confirmed by several means, including the continuum fitting method and X-ray reflection spectroscopy \cite{Bambi:2019xzp}. Moreover, the detection of a gravitational wave signal GW150914 \cite{Abbott:2016blz} by the LIGO/VIRGO Collaboration, consistently interpreted as the output of the merger of two black holes, quickly followed by the discovery of a merger of two neutron stars
GW170817 together with its optical counterpart GRB170817A \cite{TheLIGOScientific:2017qsa}, and the imaging in 2019 by the Event Horizon Telescope Collaboration of the shadow of the central object of the M87 galaxy \cite{Akiyama:2019cqa}, have further strengthen the reliability of this solution.

As we have seen in the past section, exact analytical, spherically symmetric black holes can be generated in RBGs out of (non-linear) electromagnetic fields and (anisotropic) fluids with some ease, but finding axially symmetric (rotating) black holes represent a daunting challenge for any modified theory of gravity. In this sense, the difficulty to extract exact solutions could spoil the open opportunities present to test new physics beyond GR, for instance, in the ringdown tail of gravitational waves out of binary mergers \cite{Berti:2018vdi}. To work out such scenarios within RBGs one faces a fundamental difficulty: from the structure of the field equations (\ref{eq:GmnGeneral}) and the fundamental relation (\ref{eq:qggen}) one must note that the deformation matrix ${\Omega^\alpha}_\beta$ is, in general, a nonlinear function of ${T^\mu}_\nu$, which itself depends on $g^{\mu\nu}$, while in the left-hand side of the field equations (\ref{eq:GmnGeneral}) it is $q_{\mu\nu}$ who appears instead. There are certain configurations with high symmetry (cosmology, spherically symmetric black holes, etc) in which the dependence on $g_{\mu\nu}$ can be fully removed out in favour of the matter sources, allowing to find explicit solutions using this procedure. However,  dynamical scenarios with less symmetry are plagued by technical difficulties. Moreover, the application of numerical methods on RBGs would be strongly model-dependent and computationally expensive because of the need to invert the relation between the two $q_{\mu\nu}$ and $g_{\mu\nu}$ metrics at each step. Moreover, such methods are tightly attached to the structure of Einstein's field equations, largely preventing any prospects to efficiently use their full power beyond GR.

To overcome this difficulty, an important technical progress has been recently developed and implemented, dubbed as {\bf the mapping method }\cite{Afonso:2018bpv}. It works first by introducing an Einstein frame
$G_{\mu\nu}(q)=\kappa^2 \tilde{T}_{\mu\nu}(q)$, where comparison with Eq.(\ref{eq:GmnGeneral}) yields the relation
\begin{equation} \label{eqTmunumap}
\tilde{T}{^\mu}_{\nu}(q)= \frac{1}{|\hat{\Omega}|^{1/2}}\left[{T^\mu}_\nu(g)-\delta^{\mu}_\nu\left(L_G+\tfrac{T(g)}{2}\right)\right] \ ,
\end{equation}
The new stress-energy tensor $\tilde{T}{^\mu}_{\nu}(q)$ can be derived from some new matter Lagrangian $\tilde{\mathcal{L}}_m(q_{\mu\nu},\tilde{\psi}_m)$. This establishes a correspondence  between RBGs + $\mathcal{L}_m(g_{\mu\nu},\psi_m)$ and GR + $\tilde{\mathcal{L}}_m(q_{\mu\nu},\psi_m)$, which also holds true at the level of specific solutions when supplemented with the matter field equations and the fundamental relation (\ref{eq:qggen}). To describe how this process works, let us consider the case of matter-energy sources described by anisotropic fluids, which includes a number of interesting scenarios. First, we need to write the corresponding stress-energy tensors on the RBG and GR frames as
\begin{eqnarray}
T{^\mu}_{\nu}(g)&=&\left(\rho+p_{\perp}\right) u^{\mu} u_{\nu}+p_{\perp} \delta^{\mu}_{v}+\left(p_{r}-p_{\perp}\right) \chi^{\mu} \chi_{\nu} \\
\tilde{T}{^\mu}_{\nu}(q)&=&\left(\rho^{q}+p_{\perp}^{q}\right) v^{\mu} v_{v}+p_{\perp}^{q} \delta_{v}^{\mu}+\left(p_{r}^{q}-p_{\perp}^{q}\right) \xi^{\mu} \xi_{v} \ ,
\end{eqnarray}
where $(\rho,p_r,p_{\perp})$ are the energy density, radial pressure, and tangential pressure of the fluid on the RBG frame, respectively, while $(\rho^q,p_r^q,p_{\perp}^q)$ are their counterparts on the GR frame. Plugging these expression into Eq.(\ref{eqTmunumap}) one finds the mapping equations in this case as
\begin{eqnarray}
p_{\perp}^{q} &=&\frac{1}{|\hat{\Omega}|^{1 / 2}}\left[\frac{\rho-p_{r}}{2}-\mathcal{L}_{G}\right]  \label{eq:map1} \\
\rho^{q}+p_{\perp}^{q} &=&\frac{\rho+p_{\perp}}{|\hat{\Omega}|^{1 / 2}}   \label{eq:map2}
\\ p_{r}^{q}-p_{\perp}^{q} &=&\frac{p_{r}-p_{\perp}}{|\hat{\Omega}|^{1 / 2}} \ .   \label{eq:map3}
\end{eqnarray}
These mapping equations set the following cooking recipe to produce new solutions on the RBG side out of known solutions on the GR side:

\begin{itemize}

\item Select a particular RBG coupled to some matter source described by ($\rho,p_r,p_{\perp}$) and compute ${\Omega^\mu}_{\nu}$ and $\vert \hat{\Omega} \vert$ using the fact that the latter are a function of the former.

\item Use the mapping equations (\ref{eq:map1}), (\ref{eq:map2}) and (\ref{eq:map3}) to find  ($\rho^q,p_r^q,p_{\perp}^q$) and reconstruct the matter Lagrangian on the GR side.

\item Use any known solution on GR coupled to that matter source, given by $q_{\mu\nu}$, to generate the one in RBG, $g_{\mu\nu}$, via the fundamental relation (\ref{eq:qggen}).

\end{itemize}
Let illustrate the usefulness of this program for black hole physics using two explicit examples. The first one (for purely electric fields\footnote{For an upgrade of this setting to more general classes of electromagnetic fields see Ref.\cite{Delhom:2019zrb}.}) maps GR coupled to Born-Infeld electrodynamics into EiBI gravity coupled to Maxwell electrodynamics, that is \cite{Afonso:2018mxn,Delhom:2019zrb}
\begin{eqnarray}\label{eq:corr1}
&&S_{EH} + \frac{1}{2\kappa^2\epsilon} \int d^4 x \left(1-\sqrt{1+\frac{\epsilon \kappa^2}{2\pi} X} \right) \leftrightarrow  \nonumber \\
&& \leftrightarrow  \mathcal{S}_{EiBI} + \frac{1}{8\pi} \int d^4x \sqrt{-g} X \ ,
\end{eqnarray}
where $X=-\frac{1}{2} F_{\mu\nu}F^{\mu\nu}$ is the electromagnetic field invariant. Observe how the square-root structure of Born-Infeld is transferred from the matter side to the gravity side via this correspondence. Since the corresponding black hole solutions on the GR side are known in exact form (and have been thoroughly characterized when $\epsilon<0$ \cite{Fernando:2003tz}), those of the right-hand-side of this correspondence can be worked out right away from the mapping equations (\ref{eq:map1}), (\ref{eq:map2}) and (\ref{eq:map3}), without any need to directly solving the corresponding field equations. This has been discussed in detail in Ref.\cite{Afonso:2018mxn} for the case of electrostatic field starting from the Reissner-Nordstr\"om solution of GR, showing how the hard-won solutions of Ref.\cite{Olmo:2013gqa} can be much more easily re-obtained using this procedure. This also explains why some features of the solutions obtained on the RBG side in Ref.\cite{Olmo:2013gqa} closely resemble those of the GR side.

The second example of this mapping is that GR coupled to Maxwell electrodynamics maps (surprisingly!) into EiBI gravity coupled to Born-Infeld electrodynamics, that is
\begin{eqnarray}\label{eq:corr2}
&&S_{EH} +  \frac{1}{8\pi} \int d^4x \sqrt{-g} X \leftrightarrow \nonumber \\ &&\leftrightarrow \mathcal{S}_{EiBI} +\frac{1}{2\kappa^2 \epsilon} \int d^4 x \left(1-\sqrt{1+\frac{\epsilon \kappa^2}{2\pi} X} \right) \ .
\end{eqnarray}
For electrostatic fields the solution on the left-hand side of this mapping is the Reissner-Nordstr\"om one, allowing to find via the mapping the solutions (for $\epsilon <0$) derived in Ref.\cite{Jana:2015cha} by direct calculation.

Transferring the results of any of these two mappings to the axially symmetric scenario would allow to find rotating black holes on the RBG side. In the second example (\ref{eq:corr2}), since the solution on the left-hand side of the mapping is the Kerr-Newman one of GR, one could be able to obtain the counterpart on the EiBI + BI side.
The $\epsilon$-corrections induced by this combination on the RBG side should induce qualitative and quantitative changes as compared to GR solution in terms of a modified description of horizons, ergospheres and photosphere, which would allow to test alternatives to the Kerr hyphotesis\footnote{We point out that these solutions contain an electric charge that, as we have stated despite the fact that it is negligible in astrophysical settings, in metric-affine gravity its contribution feeds the new dynamics of the solutions, being able to introduce new qualitative and quantitative differences as compared to GR solutions.} via accretion disks or different patterns in the generation of gravitational waves or in black hole shadows, all of which would presumably  be degenerated with GR predictions on  $M$, $J$, and $\epsilon$. How thus to disentangle this degeneracy in the predictions of modified gravity as compared to GR ones is still an open problem in the community. On the other hand, in the first example of the mapping above  (\ref{eq:corr1}), one could use the rotating black hole of GR coupled to Born-Infeld electrodynamics found in Ref.\cite{Toshmatov:2017zpr}\footnote{There is some discrepancy in the literature about the reliability of this solution, see the discussion of Ref.\cite{Rodrigues:2017tfm} and the alternatives provided therein.} to find the counterpart of the Kerr-Newman solution in the context of EiBI gravity, which would be regarded as a more sensible scenario from an astrophysical point of view.

The fact that the new gravitational dynamics triggered by the matter fields in metric-affine theories becomes significant in presence of high energy densities, which allow to naturally past (for most RBGs)  weak-field limit tests, also narrow the search for clear and clean observational discriminators at the typical scales of the event horizon and larger. Open opportunities can however be found in two aspects of the astrophysics of these objects. First, the modifications to the location of the photosphere would slightly change the propagation of light rays around such objects, which could be revealed via (strong) gravitational lensing, as discussed in Ref.\cite{Wei:2014dka,Sotani:2015ewa} and, more generally, could also be seen via black hole shadows \cite{Cunha:2018acu}. Second, though the propagation of gravitational waves in vacuum within RBGs are the same as in GR \cite{BeltranJimenez:2017uwv,Jana:2017ost}, their generation within binary mergers would not, suggesting the search for tiny imprints of new dynamics within the quasi-normal modes spectrum of these solutions \cite{Chen:2018vuw}. This strategy can be further reinforced by the fact that the mapping may allow also to directly implement numerical computations thanks to their mimicking of the structure of the Einstein field equations, once the correspondence between theories is identified.

\subsection{Exotic horizonless compact objects} \label{sec:scalar}

Are there any other (horizonless) compact objects rather than canonical stars?. The answer to this question is positive, and indeed a zoo of such objects can be found in the literature: gravastars, boson stars, fuzzyballs, hairy solutions, scalar clouds, gravitational solitons, etc. Many of these objects are \emph{ultra-compact}, in the sense of being close to the Buchdahl limit on compactness, namely, $\mathcal{C} \lesssim 4/9$, therefore closely resembling a black hole (which has $\mathcal{C}=1/2$) which troubles their detection via purely optical means. However, the replacement of the would-be event horizon of a black hole by a hard surface makes a fundamental difference regarding gravitational wave radiation out of binary mergers. Indeed, in one such event, besides the usual burst of gravitational waves, there will be additional modes trapped between the photosphere and the object's surface, producing a period release of secondary gravitational waves with decreasing amplitude, so-called \emph{echoes} \cite{Cardoso:2016oxy}.

Models with scalar fields offer indeed suitable avenues for the construction of such exotic compact configurations. However, despite its apparent simplicity, the resolution of the RBG field equations for scalar fields turns out to be much more harder than in the electromagnetic case due to the loss of some symmetries in the stress-energy tensor. Luckily, we have now the mapping method at our disposal. To take the simplest example of this scenario, let us consider a free, real scalar field. The mapping equations (\ref{eq:map1}), (\ref{eq:map2}) and (\ref{eq:map3}) can be conveniently applied in this case by taking into account that GR coupled to this matter source has an exact solution, studied to some detail by Wyman in Ref.\cite{Wyman:1981bd}. Therefore, if we consider quadratic $f(R)$ gravity as the target theory on the RBG side, one finds the following correspondence
\begin{eqnarray}
&&S_{EH}+\int d^4x \sqrt{-g}Z \leftrightarrow \\ \nonumber  &&   \leftrightarrow  \int d^4 x \sqrt{-g} (R-\alpha \kappa^2 R^2) + \int d^4x \sqrt{-g}(Z+\alpha \kappa^2 Z^2) \ ,
\end{eqnarray}
where $\alpha$ is a constant and $Z\equiv g^{\mu\nu}\partial_{\mu}\phi\partial_{\nu}\phi$ is the kinetic term of the scalar Lagrangian density (recall that we are taking $V(\phi)=0$). Alternatively, if we use EiBI gravity, then the mapping becomes
\begin{eqnarray}
&&S_{EH}+\int d^4x \sqrt{-g}Z \nonumber \leftrightarrow  \\&& \leftrightarrow  \mathcal{S}_{EiBI} + \frac{2}{\epsilon \kappa^2} \int d^4x \sqrt{-g}\left(\sqrt{1+\epsilon \kappa^2 Z } - 1 \right) \ .
\end{eqnarray}
From all the mappings performed so far one can see that the nonlinear structure defining either the RBG or the matter sector (either on the GR or RBG frames)  somewhat transfers from both one of the frames to the other and/or from the gravity to the matter sector (or vice-versa). Therefore, starting from  Wyman's seed solution one can generate exact solutions in the $f(R)$/EiBI gravity setting by direct algebraic transformations, as shown in Ref.\cite{Afonso:2019fzv}. The neatness of this approach is in sharp contrast with the long and cumbersome direct derivation performed in Ref.\cite{Afonso:2017aci}.

As opposed to the case of the electromagnetic fields, where the new gravitational dynamics is typically excited at the innermost region of the solutions, which is where the energy density reaches its highest values, thus leaving only tiny imprints at astrophysical scales, for scalar fields the energy density grows significantly already near the Schwarzschild radius, thus triggering a number of new properties at astrophysically-relevant scales \cite{Afonso:2019fzv}. Such properties include the presence of asymmetric wormholes, and the emergence of a kind of surface extremely close to the location of the would-be Schwarzschild horizon. As with some of their GR cousins, these exotic objects bear such a close resemblance to black holes that are hard to detect. Perhaps the best opportunity available here is also related to small differences in the generation of gravitational waves, and in the presence of echoes.

\section{Conclusions}

The physical implications of the affine connection have been mostly overlooked in the literature until very recent times, with most of the community taking it to be given by the Christoffel symbols of the metric. When its independent character is restored, it can be removed out in favour of new non-linear couplings on the matter fields (at least, in the case of minimal coupling \cite{Sotiriou:2006qn}). The corresponding metric-affine theory, when it is of RBG type, yields second-order, ghost-free, Einstein-like equations compatible with all (for most members of the family) weak-field limit and gravitational wave observations.  The resemblance of the field equations formulated this way with the standard Einstein equations allows to follow similar procedures in solving them, which has allowed to uncover a large number of theoretical results, particularly regarding exact black hole solutions.

It turns out the new gravitational contributions engendered by the matter fields in RBGs yield a number of appealing features for such solutions, which are of relevance, in particular, for the issue of singularity avoidance inside black holes. Indeed, the research carried out so far has shown that the existence of regular black holes are a resilient feature of RBGs,  including $f(R)$ gravity, quadratic gravity, and EiBI gravity, when coupled to electromagnetic fields or to different types of fluids satisfying standard energy conditions. The singularity avoidance is implemented via two different mechanisms: either through a bounce in the radial function, or by the displacement of the would-be central singularity to the future (or past) infinity of the manifold. In the former case, geodesics can naturally cross the wormhole throat, while in the latter they take an infinite time to reach to the center of the solution. Moreover, since such an avoidance turns out to be independent on the canonical scalar invariants being divergent or not, this also raises questions on the physical meaning of such scalars to characterize space-time singularities within metric-affine gravities (see Ref.\cite{Bejarano:2019zco} for a recent discussion on this point). Further extensions of RBGs, including for instance the addition to the action of scalars constructed with other contractions of the Riemann tensor requires further technical progress beyond the state-of-the-art.

When moving to realistic scenarios of interest within astrophysics, things become more involved and new strategies have to be developed. This is needed in view of the good prospects offered by these theories to testing the possible existence of new gravitational physics beyond GR within the astrophysics of compact objects. We have highlighted some of such opportunities regarding  relativistic and non-relativistic stars, black holes, and further horizonless compact objects, and some of the contributions of RBGs analyzed in the literature. Given the fact that the new gravitational dynamics in metric-affine gravities is strongly dependent on the local stress-energy densities, imprints of interest at astrophysical scales which can act as clear discriminators with respect to GR predictions are hard to find. Nonetheless, we have hinted at a few specific predictions of these theories for these objects that offer good prospects within the context of multimessenger astronomy. A  combination of such predictions for every RBG for different kinds of compact objects could allow to determine the viability of any such theory to account for different observations, thus helping to alleviate the degeneracy problem present in any modification of GR.

We have also discussed a new powerful tool to circumvent the highly non-linear character of the RBG field equations, the latter largely preventing the finding of solutions of astrophysical interest in dynamical scenarios. This tool, dubbed as the mapping method, consists on casting the RBG field equations in purely Einstenian form coupled to a new (non-canonical, in general) matter Lagrangian, in such a way that once the solution in GR for such a setting is found, the solution on the RBG side can be obtained out of it via purely algebraic transformations. The power of this method is apparent, in the sense that one can use the full machinery of analytical solutions  and numerical methods developed within GR to find new solutions on the RBG side. We have illustrated how this mapping works by discussing the way solutions found via direct resolution of the RBG field equations coupled to electromagnetic fields can be re-obtained using it. Moreover, we have discussed how new compact solutions from scalar fields can be obtained. This way of finding new solutions is absurdly simpler than the usual procedure of finding them by brute force out of the field equations, which  greatly shortens computation times and reduces the chances of mistakes. The mapping moreover allows to tackle scenarios previously unaccessible to analytic treatment, and which might be also useful for the sake of numerical simulations.

To conclude, the prospects for extracting phenomenology of interest for the physics of compact objects within metric-affine theories of gravity are exceedingly hopeful, and we cannot but to be optimistic that the field will continue to blossom in the near future.

\section*{Acknowledgements}

DRG is funded by the \emph{Atracci\'on de Talento Investigador} programme of the Comunidad de Madrid, No.2018-T1/TIC-10431, and acknowledges support from the FCT research grants Nos. SFRH/BPD/102958/2014 and PTDC/FIS-PAR/31938/2017,  the projects FIS2014-57387-C3-1-P and FIS2017-84440-C2-1-P (MINECO/FEDER, EU), the project SEJI/2017/042 (Generalitat Valenciana), and PRONEX (FAPESQ-PB/CNPQ, Brazil).  This article is based upon work from COST Actions CA15117  and CA18108, supported by COST (European Cooperation in Science and Technology). The author thanks Luis C. B. Crispino for the kind invitation to speak at the ``V Amazonian Symposium on Physics" in Bel\'em (Brazil), where this work was envisaged.



\begin{thebibliography}{00}
\bibitem{Belt}
Credit must be given to Jos\'e Beltr\'an-Jim\'enez who, to best of our understanding, was the first one in realizing this.
\end{thebibliography}
\end{document}